\documentclass{elsart}
\usepackage{t1enc,epsf}
\newcommand{\reffz}[1]{\textup{(\ref{#1})}}

\newcommand{\vect}[1]{\mbox{\boldmath $#1$}}
\newcommand{\keV}{\mbox{\rm keV}}
\newcommand{\MeV}{\mbox{\rm MeV}}
\newcommand{\GeV}{\mbox{\rm GeV}}
\newcommand{\NSt}{{\mbox{\scriptsize\rm NS}}}
\newcommand{\St}{{\mbox{\scriptsize\rm S}}}

\catcode`\@=11

\def\crrtic@{\hrule height0.1ex width0.3em}
\def\crttic@{\hrule height0.1ex width0.33em}
\def\yr{\leavevmode \kern 0.25em\raise 1.28ex\vbox{\crrtic@}\kern-0.50em}
\def\Yr{\leavevmode \kern 0.11em\raise 0.92ex\vbox{\crttic@}\kern-0.36em}
\def\yb{\leavevmode \kern 0.25em\raise 1.25ex\vbox{\crrtic@}\kern-0.55em}
\def\Yb{\leavevmode \kern 0.09em\raise 0.90ex\vbox{\crttic@}\kern-0.34em}
\def\ys{\leavevmode \kern 0.35em\raise 1.30ex\vbox{\crrtic@}\kern-0.65em}
\def\Ys{\leavevmode \kern 0.22em\raise 0.92ex\vbox{\crttic@}\kern-0.47em}
\def\yi{\leavevmode \kern 0.33em\raise 1.29ex\vbox{\crrtic@}\kern-0.63em}
\def\Yi{\leavevmode \kern 0.25em\raise 0.92ex\vbox{\crttic@}\kern-0.50em}
\def\yt{\leavevmode \kern 0.25em\raise 1.15ex\vbox{\crrtic@}\kern-0.55em}
\def\Yt{\leavevmode \kern 0.02em\raise 0.85ex\vbox{\crttic@}\kern -0.27em}
\def\yc{\leavevmode \kern -0.04em\raise 0.72ex\vbox{\crrtic@}\kern-0.26em}
\def\Yc{\leavevmode \kern 0.1em\raise  0.86ex\vbox{\crttic@}\kern-0.35em}

\def\y@{\ifcase\the\fam\yr\or\yr\or\yr\or\yr\or\yi\or\ys\or\yb
                     \or\yt\or\yc\or\ys\else\yb\fi}
\def\Y@{\ifcase\the\fam\Yr\or\Yr\or\Yr\or\Yr\or\Yi\or\Ys\or\Yb
                     \or\Yt\or\Yc\or\Ys\else\Yb\fi}
\gdef\y#1{\leavevmode\pr@kid\y@#1\pr@kid}
\gdef\Y#1{\leavevmode\pr@kid\Y@#1\pr@kid}

\def\dj{\y@ d}
\def\Dj{\Y@ D}

\begin{document}

\begin{frontmatter}
\title{Dynamical SU(3) Linear $\sigma$ Model and \\ the mixing of 
$\eta^\prime$--$\eta$ and $\sigma$--$f_0$ mesons}
\author {D. Kekez$^a$, D. Klabu\v{c}ar$^b$, M. D. Scadron$^c$}
\address{$^a$Rudjer Bo\v{s}kovi\'{c} Institute, P.O.B. 180, 
10001 Zagreb, Croatia}
\address{$^b$Physics Department, Faculty of Science,
University of Zagreb, Bijeni\v{c}ka c. 32,
Zagreb 10000, Croatia}
\address{$^c$Physics Department, University of Arizona, Tucson Az 85721 USA}

\begin{abstract}
The SU(3) linear $\sigma $ model $(L\sigma M)$ is dynamically
generated in loop-order using the nonstrange--strange basis. Only
self--consistent logarithmic divergent graphs are needed,
with quadratic divergent
graphs replaced by SU(3) mass-shell equal splitting laws. The latter
lead to an $ \eta^\prime$--$\eta $ mixing angle of $41.84^0$ which is
consistent with phenomenology. Finally this
above SU(3)$L\sigma M$ in turn predicts strong decay rates which
are all compatible with data.
\end{abstract}

\end{frontmatter}

\newpage
\section{Introduction}\label{sec:int}

The original tree--level SU(2) spontaneous symmetry breaking (SSB)
linear $\sigma$ model $(L{\sigma}M)$ interaction Lagrangian
density is \cite{Gel60}
\begin{equation}
L_{{L\sigma} M}^{\mbox{\rm int}}
=
g\bar\psi \left(\sigma+i{\gamma}_5 \vect{\tau}\cdot\vect{\pi}\right)\psi
+
g^\prime \sigma (\sigma^2+\vect{\pi}^2)
-
\frac{\lambda}{4} (\sigma^2+\vect{\pi}^2)^2~,
\label{LagDen}
\end{equation}
with tree--order chiral limiting (CL) couplings
satisfying for $f_{\pi}\approx 93~\MeV$
\begin{equation}
   g=\frac{m}{f_\pi},\,\,\,\,g^\prime=\frac{m_\sigma^2}{2f_{\pi}}
=\lambda f_{\pi}~,
\label{eq:cc}
\end{equation}
here for quark fields, constituent quark mass $m$ and cubic and
quartic--meson couplings $g^\prime$, $\lambda$. Although the $g$, $g^\prime$,
$\lambda$ in \reffz{eq:cc} are not further specified in tree order,
in loop order they are dynamically generated as
\begin{equation}
    g=\frac{2\pi}{\sqrt{3}} \approx 3.6276~,\,\,\,\,
    g^\prime = 2gm \approx 2.3~\GeV~,\,\,\,\,
    \lambda = \frac{8\pi^2}{3} \approx 26.3 \, .
\label{eq:3}
\end{equation}
We take $m=315~\MeV$, i.e., roughly one third of the nucleon mass $M_N$,
since this value
most consistently satisfies the relations central for the present 
paper, namely mass shell equal splitting laws (MSESLs) considered 
below, in Sec.~3. 
(A dynamical quark mass $m=\left[\frac{4\pi\alpha_s}{3}\langle-\bar\psi\psi
\rangle_{1\mbox{\rm\scriptsize GeV}}\right]^{1/3}\approx 320~\MeV$
used in QCD \cite{Eli84} would lead to some 3\% higher MSESLs value
$4m^2$, whereas $m=325~\MeV$, used in Refs.~\cite{Del95}, would lead to
some 6\% higher value.)
While Refs.~\cite{Del95} recover the original chiral relations of
Eqs.~\reffz{eq:cc}, the loop--order values in Eqs.~\reffz{eq:3} depend on
relating slowly diverging log--divergent graphs with the more
rapidly diverging quadratic--divergent graphs.

\hspace{4mm} In this paper we instead dynamically generate the 
SU(3) $L\sigma M$ using only the log--divergent graphs while 
replacing the quadratic--divergent graphs with the dynamical SU(3) 
MSESLs.  In Sec.~2 we review the log--divergent gap
equations in the CL and show their self--consistency in loop order.
Then in Sec.~3 we replace quadratic--divergent mass gap equations
with SU(2) and SU(3) MSESLs. In Sec.~4
we show how the SU(3) MSESLs lead to an $\eta$--$\eta^\prime$
nonstrange--strange mixing angle of $\sim 42^0$, in fact closely
agreeing with the phenomenological value \cite{Bra90,Fel98}.
Finally in Sec.~5
we employ this SU(3) $L\sigma M$ with nonstrange--strange
$\eta$, $\eta^\prime$ couplings to predict strong interaction decay rates for
$\sigma_{NS}\to\pi\pi$, $a_0\to\eta\pi$, $f_0\to\pi\pi$ and
$\eta^\prime\to\eta\pi\pi$, all in close
agreement with data \cite{Gro00}. We give our conclusions in Sec.~6.

\section{Self--consistent log divergent gap equations}\label{sec:self}

We begin with the non--perturbative loop--order equation for the
pion decay constant $\delta f_\pi =f_\pi$ in the soft--pion chiral
limit \cite{Del95}:
\begin{equation}
1
=
-i\, 4N_c\, g^2 \int \frac{\dj^{\,4}p}{ (p^{2}-m^{2})^{2} } \, ,
\label{eq:fpi}
\end{equation}
using the Goldberger--Treiman relation (GTR) $m=f_\pi g$ as in
\reffz{eq:cc} with $\dj^{\,4}p=(2\pi)^{-4} d^4 p$, where the quark mass $m$
cancels out of this gap Eq.~\reffz{eq:fpi}. This log--divergent gap
equation (LDGE) \reffz{eq:fpi} with $g\sim315~\MeV/90~\MeV \sim 3.5$
requires an ultraviolet cutoff $\Lambda\approx750~\MeV$, separating
the $\overline{q}q$ elementary particles $\pi$ and
$m_\sigma\sim 650~\MeV$ \cite{Gro00} with $m_{\pi,\sigma} < \Lambda$ from the
bound-state $\overline{q}q$ mesons $\Lambda<\rho(770)$,
$\omega(780)$, $a_1(1260)$.
This natural separation of $L\sigma M$ elementary particles from
bound states is a consequence of the $Z=0$ compositeness condition
\cite{Sal62} $g={2\pi}/\sqrt{N_c}$ or $g=3.6276$ for $N_c=3$
(also dynamically generated in Refs.~\cite{Del95}).

\hspace{4mm} The self--consistency of loops ``shrinking'' to trees 
in the CL and their link to the LDGE are seen for quark triangle and 
quark box graphs. In the former case the quark triangle representing
$g^\prime_{\sigma\pi\pi}$ is log--divergent with
\begin{equation}
g^\prime_{\sigma\pi\pi} = -8ig^3 N_c m\int{\frac{\dj^{\,4} p}{(p^2 -m^2)^2}}
=
2gm \left[ -4iN_c g^2 \int \frac{\dj^{\,4} p}{(p^2 -m^2)^2} \right]
=
2gm~,
\label{eq:5}
\end{equation}
by virtue of the LDGE \reffz{eq:fpi}. Then using the quark--level
GTR, Eq.~\reffz{eq:5} shrinks to the tree level
$g^\prime_{\sigma\pi\pi}\to g^\prime = m_\sigma^2/2f_\pi$ 
of \reffz{eq:cc} {\em provided} that $m_\sigma=2m$, the 
Nambu-Jona-Lasino \cite{Nam61} (NJL) result also dynamically generated
in Refs.~\cite{Del95}. Likewise the $\pi\pi$ box graph in the CL gives
the quartic quark coupling \cite{Del95}
\begin{equation}
    \lambda_{\mbox{\rm\scriptsize box}}
=
-8i N_c\, g^4 \int{\frac{\dj^{\,4} p}{(p^2 -m^2)^2}}
=
2g^2 \left[ -4iN_c\, g^2 \int \frac{\dj^{\,4} p}{(p^2 -m^2)^2} \right]
=
2g^2~,
\label{eq:6}
\end{equation}
again via the LDGE~\reffz{eq:fpi}. Then using the GTR, Eq.~\reffz{eq:6}
becomes
\begin{equation}
    \lambda_{\mbox{\rm\scriptsize box}}\
 = 2g^2=\frac{2g m}{f_\pi} = \frac{g^\prime}{f_\pi}
 = \lambda_{\mbox{\rm\scriptsize tree}}
\label{eq:7}
\end{equation}
by virtue of the tree--level $L\sigma M$ couplings in Eq.~\reffz{eq:cc}.
So again loops shrink to trees, while recovering the NJL scalar 
mass $m_\sigma=2m$ in this self--consistent fashion \cite{Del95}.

\hspace{4mm} Next the CL quark bubble plus quark $\sigma$ tadpole graphs,
although both being quadratically divergent give a {\em vanishing}
$m^2_\pi = 0$ in the CL (as required) provided the couplings
satisfy $g^\prime_{\sigma\pi\pi} = m_{\sigma}^2 / 2 f_\pi$, again
recovering Eq.~(\ref{eq:cc}), but independent of the quadratically
divergent scale.

\hspace{4mm} Lastly the pion quark triangle photon graph automatically
normalizes the form factor $F_{\pi} (q^2=0) = 1$ as expected.
Specifically this quark triangle predicts \cite{Pav83}
\[
F_{\pi}(q^2)
=
-i\,4N_c\, g^2 \int_0^1 dx \int \frac{\dj^{\,4}p}{[p^2-m^2+x(1-x)q^2]^2}~,
\]
recovering $F_{\pi} (q^2=0) = 1$ due to the LDGE~(\ref{eq:fpi}).

\section{Mass--Shell Equal Splitting Laws}

To continue circumventing the dangerous quadratic divergent tadpole
graphs, we first invoke the Lee null tadpole sum \cite{Lee72}
characterizing the true (not false SSB) vacuum. Using only
dimensional analysis, the vanishing tadpole sum requires \cite{Del95}
$N_{c} (2m)^4 = 3 m_\sigma^4$, or $N_c = 3$ for the SU(2)
$L\sigma M$ since we already know from Sec.~2 that the NJL
relation $m_\sigma = 2 m$ is also valid in the $L\sigma M$ in the
CL, as is $g = 2\pi / \sqrt{3}$ \cite{Sal62}.

\indent Away from the CL this NJL condition becomes for
$m \approx M_N /3 \approx 315~\MeV$,
\begin{equation}
    m^2_\sigma - m^2_\pi = 4 m^2 \approx 0.397~\GeV^2~.
\label{eq:sigma-pi}
\end{equation}
In Sec.~4 we will show that $\eta^\prime$--$\eta$ mixing requires a 
nonstrange--strange (NS--S)
mixing angle $\phi_P \approx 41.84^0$, which in turn fixes the eta
NS mass to be $m_{NS} = 757.9~\MeV$. Then the SU(3) extension of the
mass-shell equal splitting law (MSESL) Eq.~(\ref{eq:sigma-pi}) is for
$m_{a_0}\approx 984.8~\MeV$,
\begin{equation}
    m^2_{a_0} - m^2_{\eta_{NS}} \approx 0.395~\GeV^2~,
\label{eq:a0-etaNS}
\end{equation}
which we again identify with the NS quark mass factor $4m^2$ in
Eq.~(\ref{eq:sigma-pi}).

\hspace{4mm} 
MSESL (\ref{eq:a0-etaNS}) follows from the empirical $m^2_{a_0}$
and from $m^2_{\eta_{NS}}$ extracted in a phenomenological way 
in Sec.~4 (see also Ref. \cite{KeKlSc00}), but 
we have yet another way to avoid quadratic divergent amplitudes
and evaluate the difference of the $a_0$ and $\eta_{NS}$ self-energies 
explicitly. It is encouraging that this explicit calculation below
yields results which are reasonably close to MSESL (\ref{eq:a0-etaNS}), 
even though we consider only the lowest-order $L\sigma M$ self-energy 
graphs. Also, in counterdistinction 
to Eq. (\ref{eq:a0-etaNS}){\footnote{In Eq. (\ref{eq:a0-etaNS}),
we plug in $m_{NS} = 757.9~\MeV$ which {\it does} contain
\cite{KeKlSc00} the shift due to the gluon anomaly.}}, these graphs
do not capture (at least not fully) the effect of the gluon anomaly which 
influences strongly the masses in the $\eta-\eta^\prime$ complex.

\hspace{4mm} This other way to avoid quadratic divergent amplitudes is to 
subtract mass shell bubble (and tadpole) graphs of Fig.~1.
Since the quadratic divergent tadpole graphs of Figs. 1 b,d clearly cancel
(due to the $L\sigma M$ coupling relation
$g^\prime_{a_0 a_0 \sigma}=g^\prime_{\eta_{NS}\eta_{NS}\sigma}$),
the remaining quadratically divergent bubble graph difference (Figs. 1 a,c)
give the formal result
\begin{eqnarray}
m_{a_0}^2-m_{\eta_{NS}}^2
=
-8iN_c\, g^2 \int_0^1 dx \int \dj^{\,4}p
\left[ \frac{p^2 - m_{a_0}^2 x(1-x) + m^2}
     {(p^2 + m_{a_0}^2 x(1-x) - m^2)^2} \right.
\nonumber \\
-
\left. \frac{p^2 - m_{\eta_{NS}}^2 x(1-x) - m^2}
     {(p^2 + m_{\eta_{NS}}^2 x(1-x) - m^2)^2} \right]~.
\label{eq:bubble}
\end{eqnarray}

\begin{figure}[t]
\epsfxsize = 14 cm \epsfbox{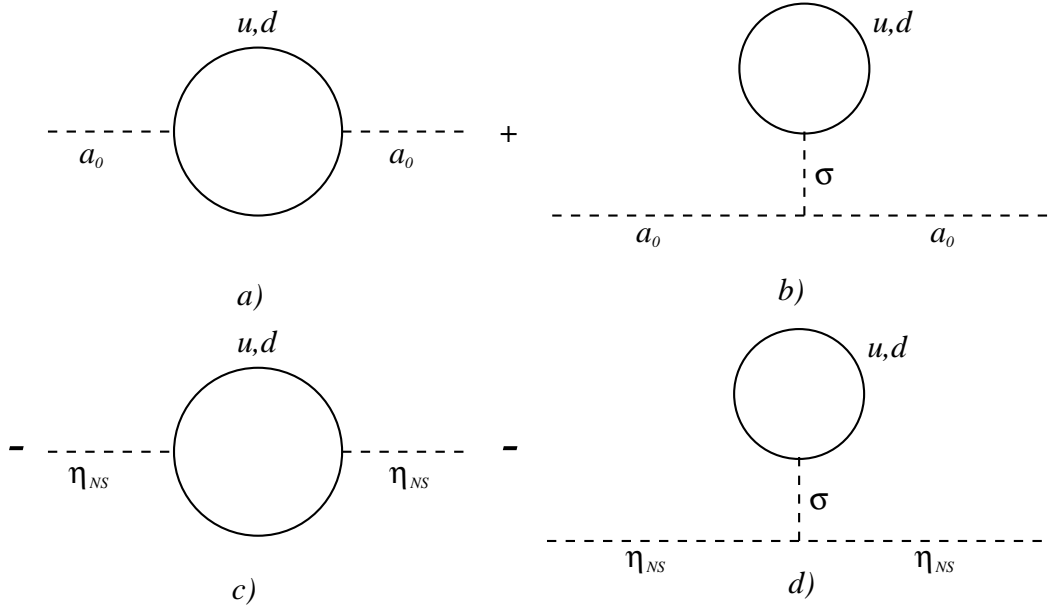} 
\caption{Bubble graphs and tadpole graphs}
\end{figure}

Here we have combined propagator denominators using Feynman's trick

\[
\frac{1}{ab}=\int_0^1 \frac{dx}{[ax+b(1-x)]^2}~.
\]

\hspace{4mm} In other words, we evaluate
the bubble graphs with propagator momentum $p\to p-qx$ for
mass shell values
$q^2 = m_{a_0}^2$, $m_{\eta_{NS}}^2$ for Figs. 1 a, c.
Specifically for $m_{a_0}=984.8~\MeV$ and $m_{\eta_{NS}}=757.9~\MeV$
and a constituent nonstrange quark mass of $315~\MeV$, a computer 
calculation detailed in the Appendix, evaluates Eq.~\reffz{eq:bubble} as

\begin{equation}
m_{a_0}^2-m_{\eta_{NS}}^2
=
5.83\, m^2 - i\, 3.83\, m^2~.
\label{eq:computer_calc}
\end{equation}

We neglect the negative imaginary part of Eq.~\reffz{eq:computer_calc} 
(compatible with unitarity)
since quarks in the bubble graph of Figs.~1 should be confined.
Considering the delicate cancellation due to the mean value
$(6\int_0^1 x(1-x)=1)$ 
giving in the second denominator in Eq.~\reffz{eq:bubble}
$m_{\eta_{NS}}^2 x(1-x)-m^2\approx -0.035\, m^2$, 
and especially recalling {\it i)} that we have used only the lowest 
order graphs and {\it ii)} that it is known that the gluon 
anomaly shifts $m_{\eta_{NS}}$ upwards [which is taken into account
in the result (\ref{eq:computer_calc}) only partially and indirectly, 
through the mass shell
value $q^2 = m_{\eta_{NS}}^2 = (757.9~\MeV)^2$] so that the difference
(\ref{eq:computer_calc}) is understandably somewhat overestimated, 
we suggest that the real part of \reffz{eq:computer_calc} is not far 
from the numerical values of Eqs.~(\ref{eq:sigma-pi}) and 
(\ref{eq:a0-etaNS}), i.e. \cite{Sca82},

\begin{equation}
m_{\sigma_{NS}}^2-m_\pi^2
=
m_{a_0}^2-m_{\eta_{NS}}^2
=
4m^2 \approx 0.397~\GeV^2~.
\label{eq:real_part}
\end{equation}

\hspace{4mm} Stated another way, combining the partial
fraction integrands in Eq.~\reffz{eq:bubble}, we
note that the leading $p^6$ terms in the numerator of \reffz{eq:bubble}
exactly cancel since they are quadratically divergent terms. Moreover,
the log--divergent $p^4$ terms in (\ref{eq:bubble}) are proportional to
$6(m_{a_0}^2-m_{\eta_{NS}}^2)x(1-x)-m^2$; they cancel using the mean value,
resulting in $m_{a_0}^2-m_{\eta_{NS}}^2=4m^2$. Thus again we support
the MSESLs of Eq.~\reffz{eq:real_part}.

\section{Dynamical $\eta^\prime$--$\eta$ Mixing}

Given the MSESL Eq.~\reffz{eq:real_part}, one may extract the NS eta
mass as $m_{\eta_{NS}}\approx 757~\MeV$. Alternatively we may express the
eta NS--S mass matrix as

\begin{equation}
\left[ \begin{array}{cc}
  m_{\eta_{NS}}^2 & \gamma \\
  \gamma & m_{\eta_{S}}^2
\end{array}
\right]
\begin{array}{c} \vspace{-4mm} \to \\ \phi_P \end{array}
\left[ \begin{array}{cc}
m_\eta^2 & 0 \\
0 & m_{\eta^\prime}^2
\end{array}
\right]~,
\label{eq:mass-matrix}
\end{equation}

where the NS--S pseudoscalar mixing angle $\phi_P$ determines
the mixing relations

\begin{equation}
|\eta\rangle = \cos\phi_P |\eta_{\mbox{\rm\scriptsize NS}}\rangle
             - \sin\phi_P |\eta_{\mbox{\rm\scriptsize S}}\rangle~,
\,\,\,\,\,\,\,
|\eta^\prime\rangle = \sin\phi_P |\eta_{\mbox{\rm\scriptsize NS}}\rangle
             + \cos\phi_P |\eta_{\mbox{\rm\scriptsize S}}\rangle~.
\label{eq:angle_def}
\end{equation}

The angle $\phi_P$ is uniquely determined via the trace constraint

\begin{equation}
m_{\eta_{NS}}^2
+
m_{\eta_{S}}^2
=
m_\eta^2 + m_{\eta^\prime}^2
\approx
1.217~\GeV^2~,
\label{eq:two-level-mixing}
\end{equation}

(because the diagonal masses $m_\eta$ and $m_{\eta^\prime}$ are measured
\cite{Gro00}) with off--diagonal hamiltonian matrix elements vanishing
$\langle\eta^\prime|H|\eta\rangle=\langle\eta|H|\eta^\prime\rangle=0$,
giving

\begin{equation}
m_{\eta_{NS}}^2
=
\cos^2\phi_P\, m_\eta^2
+
\sin^2\phi_P\, m_{\eta^\prime}^2~,
\,\,\,\,\,\,\,
m_{\eta_{S}}^2
=
\sin^2\phi_P\, m_\eta^2
+
\cos^2\phi_P\, m_{\eta^\prime}^2~.
\label{eq:m-eta-NS-S}
\end{equation}

The two--level quantum mechanical solution of
Eqs.~(\ref{eq:mass-matrix}), (\ref{eq:two-level-mixing}), (\ref{eq:m-eta-NS-S})
is the angle

\begin{equation}
\phi_P
=
\arctan\sqrt{\frac{m_{\eta_{NS}}^2-m_\eta^2}
                  {m_{\eta^\prime}^2-m_{\eta_{NS}}^2}}
=
41.84^0~,
\label{eq:phiP2}
\end{equation}

with masses

\begin{equation}
m_{\eta_{NS}} = 757.9~\MeV~,\,\,\,\,\,\,\,\,m_{\eta_S} = 801.5~\MeV~.
\label{eq:eta_NS_S_masses}
\end{equation}

\hspace{4mm} Not only is the eta NS mass in \reffz{eq:eta_NS_S_masses} close 
to $757~\MeV$ found via the MSESL Eq.~\reffz{eq:real_part}, but the mixing 
angle in \reffz{eq:phiP2} is precisely the dynamical angle obtained via
nonperturbative QCD. More specifically, Refs.~\cite{Jon79,KeKlSc00} predict
$\phi_P$ as

\begin{equation}
\phi_P
=
\arctan
\sqrt{\frac{(m_{\eta^\prime}^2-2m_K^2+m_\pi^2)(m_\eta^2-m_\pi^2)}
     {(2m_K^2-m_\pi^2-m_\eta^2)(m_{\eta^\prime}^2-m_\pi^2)}}
=
41.84^0~,
\label{eq:phiP}
\end{equation}

found from the nonperturbative QCD gluon quark annihilation strength

\begin{equation}
\beta
=
\frac{(m_{\eta^\prime}^2-m_\pi^2)(m_\eta^2-m_\pi^2)}
     {4(m_K^2-m_\pi^2)}
\approx
0.278~\GeV^2
\label{eq:beta}
\end{equation}

and a constituent quark mass ratio $X\approx 0.78\approx \hat{m}/m_s$
obtained from the NS--S QCD mass matrix \cite{Jon79,KeKlSc00}

\begin{equation}
\left[ \begin{array}{cc}
  m_\pi^2+2\beta & \sqrt{2}\beta X \\
  \sqrt{2}\beta X & 2m_K^2-m_\pi^2+\beta X^2
\end{array}
\right]
\begin{array}{c} \vspace{-4mm} \to \\ \phi_P \end{array}
\left[ \begin{array}{cc}
m_\eta^2 & 0 \\
0 & m_{\eta^\prime}^2
\end{array}
\right]~.
\label{eq:X}
\end{equation}

Equations (\ref{eq:eta_NS_S_masses}), (\ref{eq:beta}), (\ref{eq:X})
have the solution

\begin{equation}
\tan(2\phi_P)=2\sqrt{2}\beta X(m_{\eta_S}^2-m_{\eta_{NS}}^2)^{-1}=9.02
\,\,\,\mbox{\rm or}\,\,\,
\phi_P=41.84^0~.
\label{eq:tanphi_P}
\end{equation}

\hspace{4mm} Again we see that $\phi_P$ in \reffz{eq:phiP2}, \reffz{eq:phiP} and
\reffz{eq:tanphi_P} are extremely close in magnitude.
Moreover, the dynamical approach to $\eta$--$\eta^\prime$ using the
Schwinger--Dyson (SD) and Bethe--Salpeter (BS) integral equations
found $\phi_P\approx 42^0$ (that is, in terms of the singlet--octet state 
mixing  angle $\theta_P\equiv \phi_P-\arctan\sqrt{2}\approx -12.7^0$)
\cite{KlKe2}. Its subsequent refinement \cite{KeKlSc00} also included 
the effect of the ``strangeness attenuation parameter'' $X$ in the SD--BS
mass matrix. The SD--BS estimate was $X=0.663$, again close to the 
constituent quark mass ratio $\hat{m}/m_s$ found there. 
Fitting the trace constraint (\ref{eq:two-level-mixing}) then led to
$\beta = 0.277$ GeV$^2$, practically the same as Eq. (\ref{eq:beta}),
to $m_{\eta_{NS}}=757.87$ MeV and $m_{\eta_S}=801.45$ MeV, almost the same 
as Eq. (\ref{eq:eta_NS_S_masses}), and to $\theta_P=-13.4^0$, that is,
$\phi_P = 41.3^0$, very close to $\phi_P$ in Eq. (\ref{eq:tanphi_P}). 
(These results were for the original parameters of Ref. \cite{KlKe2}. 
Reference \cite{KeKlSc00} also varied the parameters to check the 
sensitivity on SD-BS modeling, but the results changed little.)
Because of the close link between
Eqs.~(\ref{eq:phiP2}), (\ref{eq:eta_NS_S_masses}), the QCD
Eqs.~(\ref{eq:phiP}), (\ref{eq:tanphi_P}) and the SD--BS scheme,
we suggest that $\phi_P\approx 41.84^0$ is the dynamical $\eta^\prime$--$\eta$
mixing angle in the NS--S quark basis. It corresponds to \cite{Bra90,Fel98}
the singlet--octet angle $\theta_P=\phi_P-\arctan\sqrt{2}\approx -12.9^0$.
Also note that Ref. \cite{KeKlSc00} showed there is no contradiction between
our approach utilizing one state--mixing angle, and the mixing scheme 
employing two angles pertaining to the mixing of the decay constants
(see Refs. \cite{Fel98}, esp. the second reference for review.) Not only is
the difference small in the NS-S basis, but our Ref \cite{KeKlSc00}
also showed that our results are in agreement with what is found in
the two--angle scheme~\cite{Fel98}.

\hspace{4mm} 
It is also satisfying that the phenomenological analysis of the NS--S
$\eta^\prime$--$\eta$ NS mixing angle extracts \cite{Bra90} 
$\phi_P = 43.2^\circ \pm 2.8^\circ$ from 
$T \rightarrow PP$ decays, $\phi_P = 36.6^\circ \pm 1.4^\circ$
from $V \rightarrow P \gamma$ and $P \rightarrow V\gamma$ decays,
$\phi_P = 41.3^\circ \pm 1.3^\circ$ from $P \rightarrow \gamma\gamma$ decays
and $\phi_P = 40.2^\circ \pm 2.8^\circ$ from $J/\psi \rightarrow \rho\eta,
\rho\eta^\prime$ and $\omega\pi^0$ decays. Moreover the recent 
Refs.~\cite{Fel98} obtain $\phi_P = 39.3^\circ \pm 1.0^\circ$  
by global phenomenological fits and $\phi_P = 42.4^\circ$ as their
theoretical prediction, which is all within the region of the dynamical 
$\phi_P$ angles in Eqs.~\reffz{eq:phiP2} or \reffz{eq:phiP}.

\section{SU(3) L$\sigma$M Strong Decay Rates}

We have thus far used the LDGE~(\ref{eq:fpi}), induced the MSESLs
$m^2_{\sigma_{NS}}-m^2_\pi=m^2_{a_0}-m^2_{\eta_{NS}} = m_\kappa^2 -
m_K^2 = 4 m^2 \approx 0.397~\GeV^2$
(for $\kappa(805-820)$ advocated by, e.g., Delbourgo and Scadron \cite{Sca82})
and the NS--S mixing angle
$\phi_P \approx 41.84^0$ all while avoiding quadratic divergent
graphs and extending the SU(2) $L\sigma M$ to SU(3). In the latter
case the cubic meson  $L\sigma M$ Lagrangian density has the SU(3)
form \cite{Sca82}

\begin{equation}
L_{\mbox{\rm\scriptsize cubic}}^{L\sigma M}
=
d_{ijk}\left( g^\prime_{SPP}S^iP^jP^k + g^\prime_{SSS}S^iS^jS^k \right)~.
\end{equation}

Then with $f_\pi \approx 93~\MeV$ and $m \approx M_N/3 \approx 315~\MeV$, 
the MSESLs above suggest the Lagrangian $g^\prime_{SPP}$ couplings

\begin{eqnarray}
g^\prime_{\sigma_{NS}\pi\pi}
&=&
\frac{m_{\sigma_{NS}\pi\pi}^2 - m_\pi^2}{2f_\pi}
\approx 2.13~\GeV~,
\\
g^\prime_{a_0 \eta_{NS} \pi}
&=&
\frac{m_{a_0}^2-m_{\eta_{NS}}^2}{2f_\pi}
\approx 2.13~\GeV~,
\\
g^\prime_{\kappa K\pi}
&=&
\frac{m_\kappa^2-m_K^2}{2f_\pi}
\approx 2.13~\GeV~,
\end{eqnarray}

along with $g^\prime_{a_0 \eta\pi}=\cos\phi_P\, g^\prime_{a_0 \eta_{NS} \pi}$,
$g^\prime_{\eta^\prime a_0 \pi}=\sin\phi_P g^\prime_{a_0 \eta_{NS} \pi}$, etc.

\hspace{4mm} 
The nonstrange $\sigma$ decay rate is predicted as \cite{Gro00,Ko94}

\begin{equation}
\Gamma(\sigma_{NS}\to\pi\pi)=\frac{3}{2}
(2 g^\prime_{\sigma_{NS}\pi\pi})^2
\frac{|\vec{p}|}{8\pi m_{\sigma_{NS}}^2}\approx 754~\MeV \, ,
\label{eq:etans-pi+pi}
\end{equation}

for $m_{\sigma_{NS}}\approx 650~\MeV$ and $|\vec{p}|=294~\MeV$. This rate is
compatible with Weinberg's mended chiral symmetry estimate \cite{Wei90}:

\begin{equation}
\Gamma_{\sigma_{NS}}
\approx
\frac{9}{2}\Gamma_\rho
\approx
676~\MeV~.
\end{equation}

\hspace{4mm} Likewise the SU(3) $L\sigma M$ $a_{0} \rightarrow \eta\pi$ decay
rate is

\begin{equation}
\Gamma_{L\sigma M}(a_0\to\eta\pi)
=
\frac{|\vec{p}|}{8\pi m_{a_0}^2}
\left[ 2 g^\prime_{a_0 \eta_{NS} \pi} \cos\phi_P \right]^2
\approx 133~\MeV
\label{eq:a0-eta+pi}
\end{equation}

for $p = 321~\MeV$, $g^\prime_{a_0 \eta_{NS} \pi} \approx 2.13~\GeV$,
$\phi_P = 41.84^0$.
One may infer a nearby $a_0$ rate from the PDG tables \cite{Gro00}.
Specifically the rate ratio 
\begin{equation}
\frac{\Gamma(a_0 \to K {\bar K})}{\Gamma(a_0 \to \eta\pi)} 
 = 0.177 \pm 0.024 
\label{rate Ratio}
\end{equation}
and $\Gamma(a_0 \to K {\bar K}) \approx 24.5$ MeV from 
Refs. \cite{OllerOset99+Astier67}, then suggests 
$\Gamma(a_0 \to \eta\pi) \approx 138$ MeV, near Eq. \reffz{eq:a0-eta+pi}.
Also, this predicted $L\sigma M$ decay rate \reffz{eq:a0-eta+pi}
is not too distant from the high statistics data \cite{Arm91}

\begin{equation}
\Gamma_{a_0 \eta\pi}
=
(95\pm 14)~\MeV~.
\end{equation}

\hspace{4mm} The SU(3) companion $f_0(980) \rightarrow \pi\pi$ rate is
estimated \cite{Gro00} to be

\begin{equation}
\Gamma(f_0\pi\pi)
\approx
(47~\MeV)(0.781)
\approx
37~\MeV
\label{eq:f0pipi}
\end{equation}

assuming the small $\Gamma(f_0 \gamma\gamma) \approx 0.56~\keV$
rate in the 1998, 1996 PDG tables combined with the measured
branching ratio $B(f_0\gamma\gamma)\approx 1.19\times10^{-5}$.
On the other hand we must account for scalar $\sigma$--$f_0$ mixing
(the analogue of pseudoscalar $\eta$--$\eta^\prime$ mixing). Thus in the
NS--S basis we define in parallel with Eq.~\reffz{eq:angle_def}

\begin{equation}
|\sigma\rangle = \cos\phi_S |\sigma_\NSt\rangle - \sin\phi_S |\sigma_\St \rangle~,
\,\,\,\,\,\,
|f_0\rangle = \sin\phi_S |\sigma_\NSt\rangle + \cos\phi_S |\sigma_\St \rangle
\, ,
\end{equation}

and estimate $\phi_S$ from the measured decay rate ratio

\begin{equation}
\frac{\Gamma(f_0\pi\pi)}{\Gamma(a_0\eta\pi)}
\approx
\frac{3}{2} \left( \frac{470~\MeV}{321~\MeV} \right)
   \left( \frac{\sin\phi_S}{\cos\phi_P} \right)^2
\approx
\frac{37~\MeV}{95~\MeV}
\approx
0.39
\,\,\,\,{\mbox{\rm or}}\,\,\,\,|\phi_S|\approx 18.3^0~.
\label{eq:BR}
\end{equation}

Prior theoretical estimates were $|\phi_s|\sim 16^0, 20^0$ \cite{Sca82}
and $14^0$ \cite{Mau99}. The DM2 data of 1989 \cite{Gro00} also suggests from
$J/\psi \rightarrow \omega\pi\pi$ that $f_0 (980)$ is mostly
$\bar{s}s$ (not nonstrange), compatible with \reffz{eq:BR} (and near the
$\phi$ (1020) which is known to be almost all $\bar{s}s$) \cite{Sca82,New20}.

\hspace{4mm} 
Lastly we calculate the strong decay rate $\eta^\prime \to \eta\pi\pi$
in the context of the SU(3) $L\sigma M$ \cite{Sch71}, with $a_0, \sigma$,
$f_0$ poles contributing as $\eta^\prime \rightarrow a_0 \pi \rightarrow
\eta \pi\pi$ (4 modes), $\eta^\prime \rightarrow \eta\sigma \rightarrow
\eta\pi\pi, \eta^\prime \rightarrow \eta f_0 \rightarrow \eta\pi\pi$.
Although the $4 a_0$ pole modes should dominate, the well--known $L\sigma M$
$\eta^\prime \rightarrow \eta \pi \pi$ contact term $3 \lambda$
{}[normalized to the quartic  term in the SU(2) Lagrangian Eq.~(\ref{LagDen})]
has the opposite sign relative to $a_0$, $\sigma$ and $f_0$ poles
and ``miraculously cancels'' them \cite{pp324} due to chiral symmetry -
assuming one treats the $a_0$, $\sigma$, $f_0$ poles in narrow width
approximation. While $\Gamma_{a_0} /m_{a_0}$, $\Gamma_{f_0}
/m_{f_0}\sim 1/10$ as needed, the $\sigma$ is {\em broad} with
$\Gamma_\sigma / m_\sigma \sim 1$.

\hspace{4mm} Then after the chiral cancellation, we must still account for the
broad--width $\sigma$ inverse propagator as $s-m^2_\sigma + i
m_\sigma \Gamma_\sigma$ with $|s - m^2_\sigma|<<|i m_\sigma
\Gamma_\sigma|$. Thus the net $\eta^\prime \rightarrow \eta{\pi^0 \pi^0}
L \sigma M$ amplitude has the magnitude

\begin{equation}
|M_{L\sigma M}^{\mbox{\rm\scriptsize net}} (\eta^\prime\to\eta\pi^0\pi^0)|
\approx
|\frac{g^\prime_{\eta^\prime\eta\sigma}g^\prime_{\sigma\pi\pi}}{m_\sigma \Gamma_\sigma}|
\approx
|\frac{g^\prime_{\eta^\prime\eta\sigma}}{2f_\pi}|
\approx
5.7~.
\label{eq:etaprime-eta+pi0+pi0}
\end{equation}

Here we \cite{KeKlSc00} estimated
$g^\prime_{\eta^\prime\eta\sigma} \approx \cos\phi_P\, \sin\phi_P\,
g^\prime_{\sigma\pi\pi} \approx 1.06~\GeV$. 
Then the net SU(3) $L\sigma M$ decay rate is predicted to be
(folding in the 3--body phase space integral \cite{Osb70})

\begin{equation}
\Gamma_{L\sigma M}(\eta^\prime\to\eta\pi^0\pi^0)
=
1.06 |M_{L\sigma M}^{\mbox{\rm\scriptsize net}}|^2~\keV
\approx
34.4~\keV~.
\label{eq:eta-eta+2pi}
\end{equation}

A slight increase of this rate \reffz{eq:eta-eta+2pi}
is due to the 10\%  non--narrow
widths of the $a_0$ and $f_0$ poles. Recent data gives \cite{Gro00}
$\Gamma (\eta^\prime \to \eta{\pi^0 \pi^0}) = (42 \pm 4)~\keV$.
The total decay rate assuming isospin invariance is

\begin{eqnarray}
\Gamma_{L\sigma M}(\eta^\prime\to\eta\pi\pi)
&\equiv&
\Gamma_{L\sigma M}(\eta^\prime\to\eta\pi^0\pi^0)
+
\Gamma_{L\sigma M}(\eta^\prime\to\eta\pi^-\pi^+)
\nonumber \\
&=&3\times(34.4\pm 4)~\keV = (103\pm 12)~\keV~,
\label{eq:etap-total}
\end{eqnarray}

near the total observed rate of
$3\times(42\pm 4)=(126\pm 12~)\keV$.
We know of no other
dynamical scheme (such as using the original singlet-octet mixing angles
\cite{Sch71}) which recover all the approximately needed SU(3) strong decay 
rates (\ref{eq:etans-pi+pi}), (\ref{eq:a0-eta+pi}), (\ref{eq:eta-eta+2pi}), 
(\ref{eq:etap-total}) as found above.

\section{Conclusion}

In this paper we have consistently avoided dealing with quadratic
divergent graphs when computing SU(2) and SU(3) linear $\sigma$
model $(L \sigma M)$ diagrams. Instead in Secs.~2 and 3 we work
only with self--consistent log--divergent gap equation 
integrals Eqs.~(\ref{eq:fpi}), (\ref{eq:bubble}). Sections 3 and 4 extend
this pattern from SU(2) to SU(3) dynamical mass-shell equal splitting laws,
leading to the off--diagonal eta nonstrange and strange
constituent quark masses $m_{\eta_{NS}}\approx 757.9~\MeV$ and
$m_{\eta_S} \approx 801.5~\MeV$. Then the dynamical $\eta^\prime$--$\eta$
mixing angle in the NS--S basis is $\phi_P \approx 41.84^0$ compatible
with nonperturbative QCD and near many phenomenological
analysis of this NS--S angle (see, e.g., Refs.~\cite{Bra90,Fel98,Jon79}).

Stated another way, the only SU(3)--breaking pattern we allow is
characterized by the constituent quark mass GTR ratio \cite{Ruj75} as used
in the phenomenological analysis of Refs.~\cite{Bra90}
$m_s / m \approx 2f_K /f_\pi - 1 \approx 1.44$
for $f_K /f_\pi \approx 1.22$ as measured
\cite{Gro00}. Then in Sec.~5 the SU(3) SPP $L\sigma M$ couplings (again
following the above MSESLs) of 
Eqs.~(\ref{eq:sigma-pi}), (\ref{eq:a0-etaNS}), (\ref{eq:computer_calc}), 
(\ref{eq:real_part})
in turn predict strong interaction $\sigma_{NS}\to\pi\pi$,
$a_0 \to \eta \pi$, $f_0 \to \pi\pi$, $\eta^\prime\to\eta\pi\pi$ 
decay rates all compatible with data \cite{Gro00}.

\vskip 4mm

{\bf Acknowledgments:}
D. Kl. and D. Ke. acknowledge the support of the Croatian Ministry of
Science and Technology under the respective contract numbers 119--222
and 009802. M. D. S. is grateful for partial support from 
the University of Zagreb and for prior conversations with A. Bramon.

\section*{Appendix: On the bubble graph integral}
                                                                                
\noindent If the integrand of Eq.~(10) is rewritten using the common
denominator ${D(x,p^2)}$, as $f(x,p^2) \equiv {N(x,p^2)}/{D(x,p^2)}$, one
should note that the $O(p^6)$ terms in its numerator $N(x,p^2)$
cancel exactly. The numerator is thus a polynomial of degree 2 in $(p^2)$:
$N(x,p^2)=c_0(x)+c_1(x)\,p^2+c_2(x)\,(p^2)^2~.$
The integrand is therefore conveniently written as the sum
 
\begin{equation}
f(x,p^2)=\sum_{i=0}^2 f_i(x,p^2)=\sum_{i=0}^2 c_i(x)\, \frac{(p^2)^i}{D(x,p^2)}~.
\end{equation}
 
\noindent The four--dimensional integral over $p$ is effectively
one-dimensional because the integrand depends on $p^2$ only.
After the Wick rotation, we performed this integration
analytically, using the {\em Mathematica} program package.
The log--divergent integral $\int d^4p\, f_2(x,p^2)$ depends on our
ultraviolet cutoff $\Lambda=750~\MeV$ required by Eq. (\ref{eq:fpi}).
After the $p^2$--integration, the logarithmic forms
 
\begin{equation}
l(x)
=
\ln\left( \frac{m^2-m_{\eta_{NS}}^2(1-x)x}{m^2-m_{a_0}^2 (1-x)x} \right)
\end{equation}
\noindent
appear in the integrand, requiring some care.                            
The mild divergences at the points
$x_0=0.115698$, $x_1=0.222046$, $x_2=0.777954$, and $x_3=0.884302$
correspond to the
roots of polynomials $x\mapsto m^2-m_{\eta_{NS}}^2(1-x)x $
and $x\mapsto m^2-m_{a_0}^2 (1-x)x$.
In order to perform the residual $x$ integration
of the functions $x\mapsto \int d^4p\,f_i(x,p^2)$ $(i=0,1,2)$,
the interval $[0,1]$ is divided into five integration regions,
$[0,x_1]$, $[x_1,x_2]$, $[x_2,x_3]$, $[x_3,x_4]$, and $[x_4,1]$.
These integrations were numerical, with an
adaptive algorithm
which can handle the mild, integrable singularities appearing at the
edges of the integration regions.

\end{document}